\documentclass[aps,pra,pdf,superscriptaddress,amsmath,amssymb,amsfonts,twocolumn,showpacs,nofootinbib,longbibliography]{revtex4-1}

\usepackage{silence}
\usepackage{amssymb}
\usepackage{eqnarray,amsmath}
\usepackage{dcolumn}
\usepackage{graphicx}
\usepackage{natbib}
\usepackage{bbm}
\usepackage{amssymb,amsmath}
\usepackage{latexsym}
\usepackage{amsfonts}
\usepackage{epsfig}
\usepackage{psfrag}
\usepackage{graphicx}
\usepackage{hyperref}
\usepackage{color}
\usepackage{soul,xcolor}
\usepackage[normalem]{ulem}
\usepackage[titletoc]{appendix}
\usepackage{float}
\usepackage{dcolumn}
\usepackage{bm}
\usepackage{braket}
\usepackage{amsmath}
\usepackage{physics}
\usepackage{mathtools}
\usepackage{graphicx,color,xcolor,colortbl}

\newcommand{\hide}[1]{}

\begin{document}

\title{Engineering chiral spin interactions with Rydberg atoms}

\author{Elena Kuznetsova}
\thanks{The work began at ITAMP at Harvard University.}
\address{ITAMP, Center for Astrophysics $|$ Harvard $\&$ Smithsonian, 60 Garden Street, Cambridge, Massachusetts 02138, USA}
\address{Department of Physics, Harvard University, Cambridge, Massachusetts 02138, USA}
\author{S. I. Mistakidis}
\address{ITAMP, Center for Astrophysics $|$ Harvard $\&$ Smithsonian, 60 Garden Street, Cambridge, Massachusetts 02138, USA}
\address{Department of Physics, Harvard University, Cambridge, Massachusetts 02138, USA}
\author{Seth T. Rittenhouse} 
\address{Department of Physics, the United States Naval Academy, Annapolis, Maryland 21402, USA}
\address{ITAMP, Center for Astrophysics $|$ Harvard $\&$ Smithsonian, 60 Garden Street, Cambridge, Massachusetts 02138, USA}
\author{Susanne F. Yelin}
\address{Department of Physics, Harvard University, Cambridge, Massachusetts 02138, USA}
\author{H. R. Sadeghpour}
\email{Corresponding author: hrs@cfa.harvard.edu}
\address{ITAMP, Center for Astrophysics $|$ Harvard $\&$ Smithsonian, 60 Garden Street, Cambridge, Massachusetts 02138, USA}

\date{\today}

\begin{abstract}
We propose to simulate the anisotropic and chiral Dzyaloshinskii-Moriya (DM) interaction with Rydberg atom arrays.
The DM Hamiltonian is engineered in a one-dimensional optical lattice or trap array with effective long-range Rydberg spins, interacting indirectly via a mobile mediator Rydberg atom. A host of XXZ and DM Hamiltonians can be simulated with out-of-phase sign periodic coupling strengths; for initial states in a stationary condensate, the DM interaction vanishes. 
This theory allows for determination of the DM interaction (DMI) vector components from first principles. The inherent anisotropy of the Rydberg-Rydberg interactions, facilitates the DMI coupling to be tuned so as to be comparable to the XXZ interaction. 
Our results make plausible the formation of non-trivial topological spin textures with Rydberg atom arrays.    

\end{abstract}

\maketitle

{{\it Introduction.} The versatility and scalability of quantum simulators, made with superconducting qubits~\cite{arute2019quantum}, trapped ion arrays~\cite{zhang2017observation}, neutral atoms in optical lattices~\cite{bloch12} and tweezers~\cite{kaufman2021quantum}, render a range of complex quantum many-body problems, solvable. Quantum magnetization mainly appears in the symmetric flavor, the (${\bf S}_i\cdot {\bf S}_j$) spin-spin interactions, as in the Ising or Heisenberg magnetization. These interacting spin Hamiltonians are prototypical models for simulating complex quantum many-body, often correlated, condensed or solid state matter systems~\cite{QSim-review}. Cold atomic and molecular systems offer ideal platforms to emulate and control large interacting spin Hamiltonians ~\cite{yan12,hild2014far,jepsen2020spin} 
describing quantum magnetic setups. 

The less commonly considered anti-symmetric spin-spin interaction, namely the (${\bf S}_i\times {\bf S}_j$) term, on the other hand, leads to non-trivial topological textures, such as domain walls~\cite{DW1,DW2,DW3,kuepferling2023measuring}, magnetic skyrmions~\cite{skyrmion,skyrmion2}, hopfions~\cite{hopfion}, and topological magnons~\cite{magnons}. Originally proposed by Dzyaloshinskii and Moriya~\cite{DMI1,DMI2}, to explain the appearance of weak ferromagnetization in antiferromagnets, the eponymous Hamiltonian, $H_{ {\rm DM}}=\Sigma_{i,j} {\bf D_{ij}}\cdot ({\bf S_i}\times {\bf S_j})$, where ${\bf D_{ij}}$ is the vector coupling strength between spins ${\bf i}$ and ${\bf j}$, breaks the inversion symmetry in a magnetic system. The discovery of magnetic skyrmions in chiral magnets~\cite{skyrmion3} has led to new appreciation for the role played by the DMI in topological magnetic textures. A technical review of the literature and methods to calculate and simulate DMI in condensed matter and spintronics can be found in Ref.~\cite{DM3}. However, the chiral DM Hamiltonian has yet to be simulated in an atomic molecular and optical physics setting.

In this letter, we propose a scheme based on Rydberg atom platforms to realize the chiral DM Hamiltonian. 
It is demonstrated that Rydberg atoms in tweezer arrays, interacting via tunable dipole-dipole coupling, with a mobile mediator Rydberg atom in a bilayer setup, generate a complex exchange coupling. 
Its real component is related to the usual Heisenberg exchange and the complex contribution is proportional to the DMI~\cite{shekhtman1992moriya}. 
An interesting corollary of our Rydberg excitation proposal is that the DMI vector coefficients can be calculated from first principles. In addition, precise tuning of DMI vs. the Heisenberg exchange interaction (J) is possible both in the weak ($|{\bf D_{ij}}|/J <1$) and strong ($|{\bf D_{ij}}|/J \gg 1$) coupling regimes thus rendering formation of chiral topological spin textures possible.

The DMI belongs to a particular class of indirect, i.e., mediated, spin-spin interactions. Indirect interactions such as the phonon-mediated electron-electron~\cite{Phonon-mediated-electron-inter} and Ruderman-Kittel-Kasuya-Yosida (RKKY)~\cite{RKKY-our-PRA} can be simulated in bilayer Rydberg atom setups. First steps towards simulating the DM Hamiltonian with Rydberg atoms have only recently been discussed~\cite{kunimi2023,nishad2023}. The proposal by Kunimi {\it et al.}~\cite{kunimi2023} exploits the rotation angle between a Raman laser and an array of Rydberg atoms to engineer the DMI. 
In Nishad {\it et al.}~\cite{nishad2023}, the XY Heisenberg exchange picks up a DMI term by modulating a one-dimensional chain of 4-atom segments.

Real materials displaying DMI, such as engineered chains of magnetic Fe, Co, and Mn atoms adsorbed 
on metal surfaces~\cite{DMI-Weisendanger1,DMI-Weisendanger2,DMI-Weisendanger3,DMI-Weisendanger4,DMI-Weisendanger5,DMI-Weisendanger6,shen2022effects}, allow some parameters to be tuned by geometric tailoring~\cite{DMI-geometry-tuning} and external electric fields~\cite{DMI-electric-field-tuning}. They are, however, by far exceeded by the tunability available in ultracold atomic setups, in which dimensionality, geometry, disorder, connectivity and strength of 
effective spin-spin interaction 
can be precisely controlled. 
Indeed, ultracold trapped neutral atoms and polar molecules, or trapped ions are particularly suitable for quantum simulation of magnetic models~\cite{QSim-review}, since they often possess long-lived states for encoding effective spins and can interact via long-range interactions. 
Moreover, cold atoms and molecules, placed in periodic tweezer arrays, can scale up to $\sim$1000 spins~\cite{2D-Ising-Rydbergs1,2D-Ising-Rydbergs2,Rydberg-arrays1,Rydberg-arrays2,Rydberg-arrays3} for large-scale simulations  
~\cite{QSim-ultracold-atoms-review}.

{\it Bilayer approach.}
Here, we consider a bilayer system, consisting of an  
array of Rydberg atoms encoding effective spins in atomic Rydberg states. 
They interact via dipole-dipole with mediator Rydberg atoms moving in parallel to the array. 
We demonstrate that this arrangement gives rise to DMI between the effective spins. The requisite condition necessary for the realization of DMI~\cite{DMI}, i. e. breaking the inversion symmetry, is satisfied by preparing a mediator Rydberg atom with  nonzero quasimomentum. 

\begin{figure}[ht!]
\center{\includegraphics[width=1.0\linewidth]{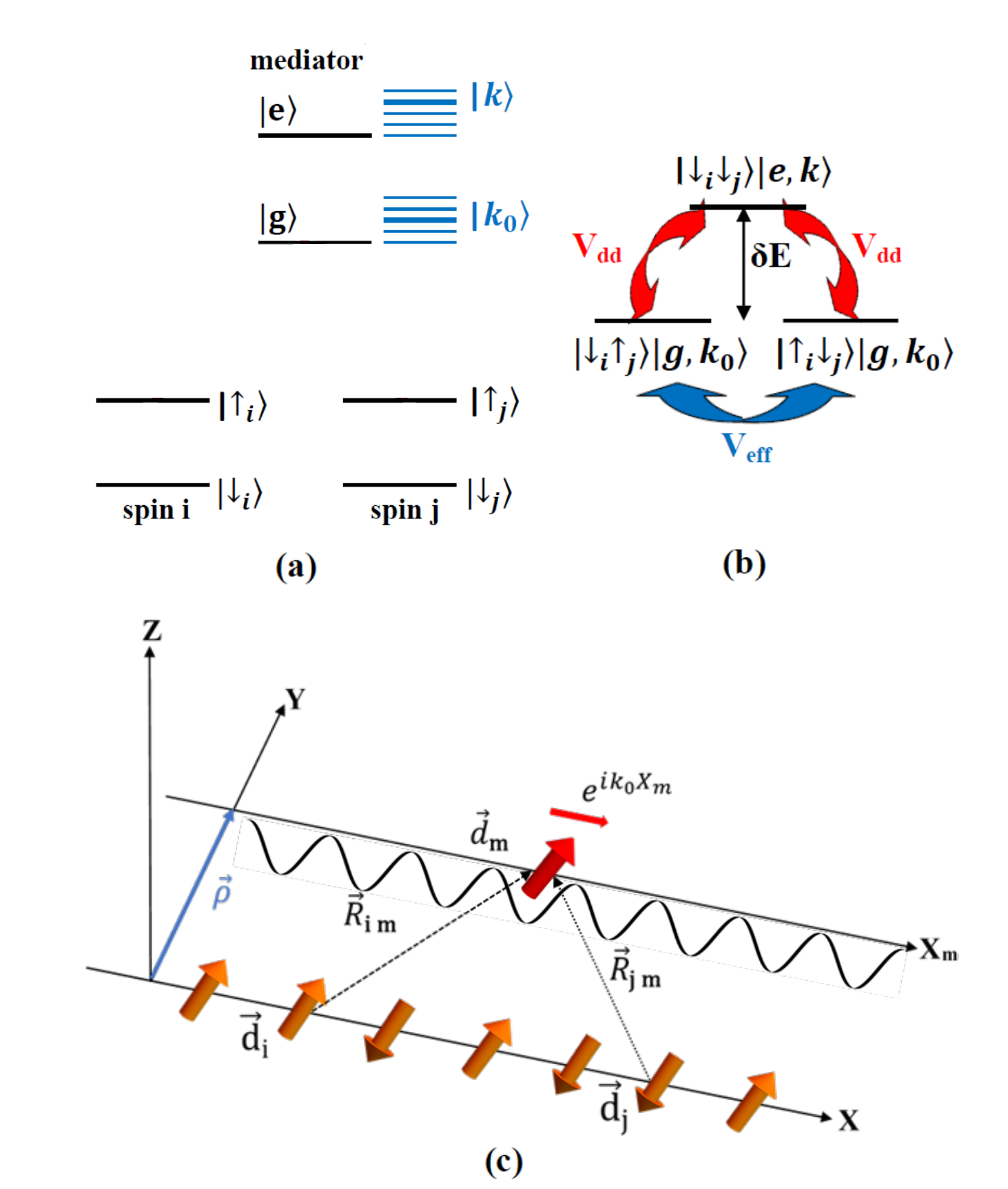}
\caption{\label{fig:simplified-setup} {\it Schematic representation of the bilayer Rydberg setup and the involved spin states.} (a) Effective i$^{\rm th}$ and j$^{\rm th}$ spin-1/2 particles and a two-level mediator which can occupy nonzero quasimomentum states e.g.  $\ket{k_0}$, $\ket{k}$ in the ground and excited states respectively.  
(b) For $|\hat{V}_{dd}| \ll \delta E$ the effective interaction between $\ket{\downarrow_{i}\uparrow_{j}}\ket{g,k_{0}} \leftrightarrow \ket{\uparrow_{i}\downarrow_{j}}\ket{g,k_{0}}$ arises within second-order perturbation theory via virtual states $\ket{\downarrow_{i}\downarrow_{j}}\ket{e,k}$. The $\ket{g, k_0}$ and $\ket{e,k}$ refer to ground and excited mediator internal and motional states respectively. 
(c) Bilayer setting, where effective spins with dipole moments ${\bf d}_{i}$ and ${\bf d}_{j}$ interact with a mediator dipole ${\bf d}_{\rm m}$. 
The mediator moves parallel to the spin array in a shallow optical lattice with initial quasimomentum $k_{0}$. The distance between the arrays is ${\bf \rho}$, and ${\bf R}_{i \rm m}$ is the separation vector between the i$^{\rm th}$ spin and the mediator.}}
\end{figure}

The spins can be in one of two states, $\ket{\uparrow}$ or $\ket{\downarrow}$ and are trapped in a deep optical lattice, while the mediator experiences a shallow lattice and it is prepared in an internal state $\ket{g}$ with initial quasimomentum $k_{0}$. 
The spin and the mediator interact via dipole-dipole coupling which modifies the internal spin states and both 
internal and motional states of the mediator, see Fig.~\ref{fig:simplified-setup}(a). 
We consider dipolar strengths much smaller than the energy
difference between the mediator and spin transitions, i.e.  $|\hat{V}_{dd}| \ll \delta E$, where 
$\delta E=E_{e}+E_{\rm kin}(k)-(E_{g}+E_{\rm kin}(k_{0}))-(E_{\uparrow}-E_{\downarrow})$. 
Here, $E_{g}$ ($E_{e}$) refer to the energy of the ground (excited) mediator state, $E_{\downarrow}$ ($E_{\uparrow}$) represents the total energy of the spin-$\downarrow$ ($\uparrow$) and $E_{\rm kin}(k_{0})$ ($E_{\rm kin}(k$) denotes the kinetic energy of the mediator with quasimomentum $k_0$ ($k$).  
Under these conditions, transitions involving simultaneous spin-flips and excitation of mediator states such as $\ket{\uparrow_{i}}\ket{g,k_{0}} \leftrightarrow \ket{\downarrow_{i}}\ket{e,k}$ are included only virtually.  

The effective spin-spin interaction within second-order perturbation theory [Fig.\ref{fig:simplified-setup}(b)] qualitatively takes the form
\begin{align}
\label{eq:Veff-simplified}
\hat{V}_{\rm eff}\sim \sum_{k}\frac{
\hat{V}_{dd}\ket{e,k}\bra{e,k}\hat{V}_{dd}^{\dagger}
}{\delta E}, 
\end{align}
where the dipolar interaction among the $N$ spins is given by   
\begin{align}
\label{eq:dip-dip-inter-init}
\hat{V}_{dd}=\sum_{i=1}^{N}\frac{\hat{{\bf d}}_{i}\cdot\hat{{\bf d}}_{\rm m}}{R_{i\rm m}^{3}}&-\frac{3\left(\hat{{\bf d}}_{i}\cdot{\bf R}_{i \rm m}\right)\left(\hat{{\bf d}}_{\rm m}\cdot{\bf R}_{i \rm m}\right)}{R_{i \rm m}^{5}}  \nonumber \\
 &\sim \sum_{i} \mathcal{V}({\bf R}_{i \rm m})\hat{S}_{i}^{+}\hat{\sigma}_{\rm m}^{-} + {\rm H.c.}.
\end{align}
Here, ${\bf R}_{i \rm m}={\bf R}_{i}-{\bf R}_{\rm m}$ with ${\bf R}_{i}$ (${\bf R}_{\rm m}$) denoting the position of the i$^{\rm th}$ spin (mediator) and $\hat{{\bf d}}_{i}=\hat{S}^{+}_{i}{\bf d}^{\uparrow \downarrow}_{\rm spin}+{\rm H.c.}$  ($\hat{{\bf d}}_{\rm m}=\hat{\sigma}^{+}_{\rm m}{\bf d}^{eg}_{\rm m}+{\rm H.c.}$) 
is the i$^{\rm th}$ spin (mediator) electric dipole operator. 
Moreover,  ${\bf d}^{\uparrow \downarrow}_{\rm spin}$ (${\bf d}^{eg}_{\rm m}$) is the dipole matrix element between the $\ket{\uparrow}$ and $\ket{\downarrow}$ spin ($\ket{g}$ and $\ket{e}$ mediator) states. 
Also, $\hat{S}^{+}_{i}=\ket{\uparrow_{i}}\bra{\downarrow_{i}}$ and 
$\hat{\sigma}^{+}_{\rm m}=\ket{e}\bra{g}$, refer to the 
effective spin and mediator raising operators, while  $\mathcal{V}({\bf R}_{\rm im})$ is the position dependent part of the dipole-dipole interaction. 

We highlight the key idea here: by considering the asymptotic form of the dipolar interaction   
$\mathcal{V}({\bf R}_{i \rm m}) \sim \mathcal{V}_{dd}/{\bf R}_{i \rm m}^3$, and taking the wave functions of the mediator motional states as plane waves $\bra{{\bf R}_{\rm m}}k\Big\rangle \sim e^{ikX_{\rm m}}$, 
the interaction matrix elements become 
\begin{align}
\bra{k_{0}} \mathcal{V}({\bf R}_{i \rm m})\ket{k} \sim  \mathcal{V}_{dd}e^{i(k-k_{0})X_{i}}, \nonumber
\end{align}
where $X_{i}$ is the position of the i$^{\rm th}$ spin along the $X_{\rm m}$ direction in which the mediator is set to move. A rigorous derivation is provided in the supplemental material (SM)~\cite{supp}.

\begin{figure}
\center{\includegraphics[width=8.5cm]{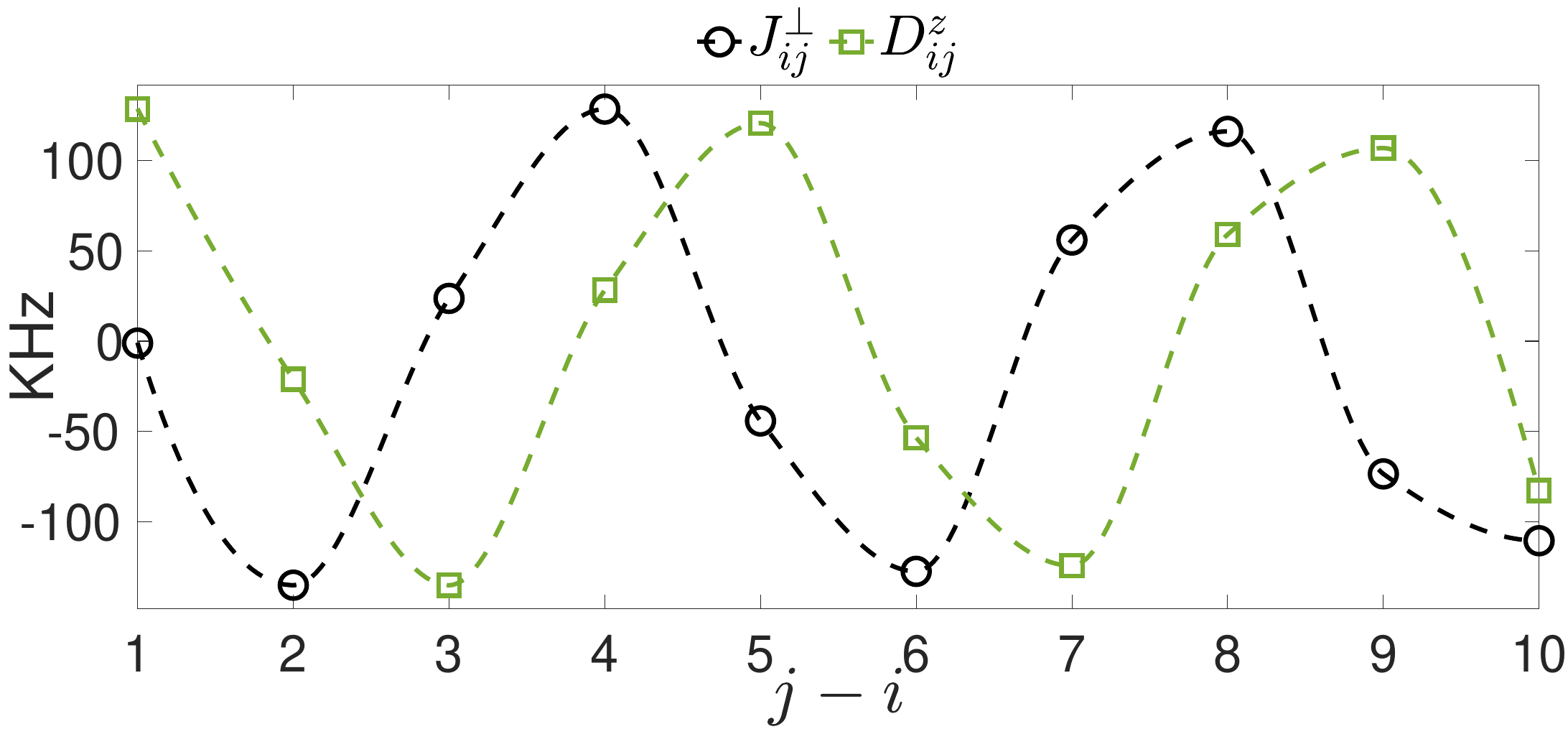}
\caption{\label{fig:J-D-graphs} {\it Out-of-phase sign periodic  behavior of the XX and DM interaction coefficients.} (a) $J_{ij}^{\bot}$ and (b) $D_{ij}^{z}$ coefficients numerically obtained via Eq.~(\ref{int_coef_approx}) in the lowest Bloch band with $N_{\rm spin}=100$, $N_{\rm m}=100$, $a_{\rm spin}=a_{\rm m}=10$ $\mu$m, $\rho=2$ $\mu$m, $k_{0}=48\pi/L_{\rm m}$ and mediator's lattice depth $V_{0}=-E_{R}$. Dashed lines guide the eye.}} 
\end{figure}

Importantly, the effective interaction of Eq.~(\ref{eq:Veff-simplified}) takes the form 
\begin{align}
\label{eq:Veff-approx}
\hat{V}_{\rm eff} \sim \sum_{i,j}J_{ij}^{\pm}\hat{S}_{i}^{+}\hat{S}_{j}^{-} + {\rm H.c.},
\end{align}
with the coefficients 
\begin{align}
J_{ij}^{\pm} \sim \sum_{k}\frac{| \mathcal{ V}_{dd}|^{2}e^{i(k-k_{0})(X_{i}-X_{j})}}{\delta E} \sim e^{-ik_{0}(X_{i}-X_{j})}. \label{int_coef_approx}
\end{align}
The summation over quasimomenta is in the first Brillouin zone.
For $k_{0} \neq 0$ the coefficients $J_{ij}^{\pm}$ are in general complex. 
With some re-arrangements the effective Hamiltonian becomes 
\begin{align}
\hat{V}_{\rm eff} \sim &{\rm Re}J_{ij}^{\pm}\left(\hat{S}_{i}^{+}\hat{S}_{j}^{-}+\hat{S}_{i}^{-}\hat{S}_{j}^{+}\right)+
i{\rm Im}J_{ij}^{\pm}\left(\hat{S}_{i}^{+}\hat{S}_{j}^{-}-\hat{S}_{i}^{-}\hat{S}_{j}^{+}\right)  \nonumber \\
&=\frac{J_{ij}^{\bot}}{2}\left(\hat{S}_{i}^{+}\hat{S}_{j}^{-}+\hat{S}_{i}^{-}\hat{S}_{j}^{+}\right)+
\bold{D}_{ij} \cdot \left(\hat{\bold{S}}_{i}\times \hat{\bold{S}}_{j}\right)\label{effective_pot}
\end{align}
where the coefficients of the XX and DM interactions 
\begin{align}
J_{ij}^{\bot}=2{\rm Re}J_{ij}^{\pm},~~~\textrm{and}~~~\bold{D}_{ij}=\left(0,0,2{\rm Im}J_{ij}^{\pm}\right).\label{coefficientsXX_DM}
\end{align} 
Therefore, the effective Hamiltonian is of XX type (XXZ in the general case~\cite{alcaraz1990heisenberg}) 
with an additional DMI term $\bold{D}_{i,j} \cdot \left(\hat{\bold{S}}_{i}\times \hat{\bold{S}}_{j}\right)$. 
It is the complex nature of $J_{ij}^{\pm}$ due to the modification of the mediator  motional states induced by the spin-mediator interaction and its 
non-zero quasimomentum ($k_{0} \neq 0$) that give rise to the effective DMI.}

From the above Eqs.~\ref{int_coef_approx}- \ref{coefficientsXX_DM}, the XX and DM coefficients  behave as $J_{ij}^{\bot} \sim (-1)^{|i-j|}\cos \big[ k_{0}(X_{i}-X_{j}) \big]$ and $D_{ij}^{z} \sim (-1)^{|i-j|+1}\sin \big[ k_{0}(X_{i}-X_{j}) \big]$. They are out-of-phase sign periodic and of infinite range. Upon considering an initial mediator state with a Lorentzian momentum distribution, $p_{k_{s}}=(\delta k_{0}/\pi)/((k_{s}-k_{0})^{2}+\delta k_{0}^{2})$, then the averaged interaction coefficients have a decay profile,  $J_{ij}^{+-} \sim e^{-\delta k_{0}|X_{i}-X_{j}|} e^{-ik_{0}|X_{i}-X_{j}|} $. 

When using Rydberg states as spins, the direct spin-spin interactions can be explicitly calculated and are generally large compared to the $V_{\rm eff}\sim$100 kHz expected here.  However, dipole-dipole interactions are anisotropic with an angular dependence $P_2(\cos \theta)$ (where $P_2$ is a second order Legendre polynomial) meaning that they may be tuned by changing the excitation laser polarization angle with respect to the lattice direction. Choosing the so called ``magic'' angle of $\theta=54.7^\circ$, fully suppresses the direct interactions.  

The above discussion gives a qualitative description of the emergence of the asymmetric DM interaction.  A rigorous derivation of $\hat{V}_{\rm eff}$ at the level of second order perturbation theory is given in the SM~\cite{supp}. 
The results from this rigorous treatment are used throughout the remainder of this letter.

{{\it Implementation.}
A realistic setup to extract the XX and DMI  coefficients consists of an array of $^{87}$Rb Rydberg atoms interacting with a Rydberg mediator moving in a parallel lattice. It is possible to induce strong interactions via a F\"{o}rster resonance~\cite{Rb-spin-mediator-encoding}. 
The spin and mediator states can be 
encoded in $\ket{\uparrow_{i}}=\ket{48p_{1/2}}$, $\ket{\downarrow_{i}}=\ket{48s_{1/2}}$, and 
$\ket{e}=\ket{50s_{1/2}}$, $\ket{g}=\ket{49p_{1/2}}$ respectively.  
The F\"{o}rster defect between the $\ket{\uparrow_{i}} \to \ket{\downarrow_{i}}$ and $\ket{e} \to \ket{g}$ in  $^{87}$Rb is  
$\Delta E \approx 117$MHz~\cite{Rb-50s-49p-encoding}. The lowest magentic sublevels $m_j =1/2$ states can be tuned to resonance with the application of a magnetic field of $B\approx 58$ G, see SM~\cite{supp}. 
  
The mobile mediator in the lowest Bloch band is initialized via an excitation $\ket{5s_{1/2}} \to \ket{49p_{1/2}}$, utilizing a three photon excitation scheme which allows for controlling the preparation of the momentum $k_0$ of the mediator as discussed in the SM~\cite{supp}. 
Such a recoil-less, Doppler-free excitation of Rb($np$) states was recently demonstrated in Ref.~\cite{Doppler-recoiless-Rydb-excit}. 
An adiabatic ramp of the lattice in which the mediator lies ensures that $k_{0}$ remains intact and the mediator resides in the lowest Bloch band. 
The mediator motional states, $\bra{X_{\rm m}}k\rangle =u_{k}^{(\nu)}(X_{\rm m})e^{ikX_{\rm m}}/\sqrt{L_{\rm m}}$, are characterized by quasimomentum $k$ in the $\nu$ Bloch band. 
The mediator lattice of length $L_{\rm m}=a_{\rm m}N_{\rm m}$ with $N_{\rm m}$ sites and lattice constant $a_{\rm m}$ [Fig.\ref{fig:simplified-setup}(c)] is subject to periodic boundary conditions. 

The XX and DM interaction strengths are explicitly calculated in the SM~\cite{supp}, see for instance Eq.~(\ref{eq:J-coeff-main}). 

\begin{equation}
\label{eq:J-coeff-main}
\begin{split}
J_{ij}^{+-}=\frac{1}{2}&\sum_{k_{\nu}}\bigg( \frac{V_{k_{\nu_0}k_{\nu},i}^{\uparrow \downarrow,ge}V_{k_{\nu}k_{\nu_0},j}^{\downarrow \uparrow,eg}}{E_{\rm spin}+E_{\rm kin}(k_{\nu_{0}})-E_{\rm m}-E_{\rm kin}(k_{\nu})} \\&-\frac{V_{k_{\nu_0}k_{\nu},i}^{\downarrow \uparrow,ge}V_{k_{\nu}k_{\nu_0},j}^{\uparrow \downarrow,eg}}{E_{\rm spin}+E_{\rm m}+E_{\rm kin}(k'_{\nu'})-E_{\rm kin}(k_{\nu_{0}})} \bigg),
\end{split}
\end{equation}
where $V_{k_{\nu},k'_{\nu'},i}^{ \alpha \beta,\eta \xi}$ represent the dipole interaction matrix elements for the $i^{\textrm{th}}$ spin, while $\alpha\beta=\uparrow \downarrow,\downarrow \uparrow$, and $\eta \xi=eg,ge$. 
The $J_{ij}^{\bot}$ and $D_{ij}^{z}$ coefficients, shown in Fig.~\ref{fig:J-D-graphs}, exhibit an out-of-phase sign periodic behavior, where the DMI dominates for odd interspin separations, while the XX interaction is the strongest 
for even interspin separations.  
For the parameters used in Fig.~\ref{fig:J-D-graphs} it holds that $k_{0}\approx \pi/2a_{\textrm m}$. Therefore, $J_{ij}^{\bot} \sim \cos \big[ \pi(i-j)/2 \big]$, $D_{ij}^{z} \sim \sin \big[ \pi(i-j)/2 \big]$, which agree qualitatively with the results in Fig.~\ref{fig:J-D-graphs}. This qualitative argument suggests that the relative strengths of the DM and XX interactions may be tuned by varying $k_{0}$ as $|D_{ij}^{z}/J_{ij}^{\bot}| \sim |\tan \big[ k_{0}(X_{i}-X_{j}) \big]|$. In particular, for $k_{0}=0$, corresponding 
to the mediator prepared initially in a stationary Bose condensate, the DMI vanishes~\cite{RKKY-our-PRA}, and only sign changing XX interaction of 
RKKY type is present. 
The cycle time for both XX and DM interactions are $2\pi/|J_{ij}^{\bot}|,\ 2\pi/|D_{ij}^{z}| \approx 8$ $\mu$s, which allows for $\sim 20-30$ coherent interactions in a typical Rydberg lifetime.

{\it Summary and Outlook.} We demonstrate that  chiral DM interactions can be realized in  bilayer Rydberg array platforms with mobile mediator Rydberg atoms. An array of effective 
spins couple via dipolar interactions with a two-level Rydberg mediator moving in a parallel optical lattice. 
The strength of the XXZ and DM coefficients can be controlled with the mediator initial quasimomentum. A unique aspect of the proposal lies in the fact that the XXZ and DM interactions with Rydberg atoms can be calculated from first principles. 
We note that the above-described scheme can also be implemented with an array of polar molecules encoding spins in rotational states~\cite{RKKY-our-PRA}.

For the DM and XXZ interactions of comparable strength, non-collinear spin ground states, topologically non-trivial states- domain walls, skyrmions  and magnons- as well as spin textures with position dependent chiralty can be realized. Both the XXZ and DM interactions can extend beyond nearest neighbors and are shown to be out-of-phase sign periodic on the interspin separation with finite range. The long-range sign-changing XXZ interaction analogous to the RKKY ~\cite{RKKY1,RKKY1_1,RKKY2,RKKY3,RKKY4,RKKY5,RKKY6,RKKY7,RKKY8}, leads to collinear spin orderings, while the 
DMI part results to non-collinear spin textures with position 
dependent chirality~\cite{DMI-RKKY1,DMI-RKKY2,DMI-RKKY3,DMI-RKKY4}. 
The chiral nature of the DMI can be exploited to create arbitrary spin textures and chiral magnetic phenomena~\cite{1D-chain-DM-theory_1,1D-chain-DM-theory_2,1D-chain-DM-theory_3,1D-chain-DM-theory_4,1D-chain-DM-theory_5,skyrmion2}. Most applications of DMI in interfacial materials involve weak chirality, i.e. $|D^{z}/J^{\bot}|\ll 1$. Here, we show that strong DMI regime can be realized with Rydberg interactions.

It is also possible to engineer chiral DM interactions with ground state atoms in a ring geometry, interacting with a central spin Rydberg atom with strong spin-orbit interaction, such as a Cs($nd$) atom, see Sec.~\ref{IV} in SM for details~\cite{supp}. Under this scenario, excitations perpendicular to the plane of the ring with large magnetic quantum numbers create electronic lobes, so that the spin-orbit coupling can break the spin inversion symmetry. Impurity electron-spin interactions in the Friedel virtual bound states model of the scattering of conduction electrons from localized impurities is reminiscent of the Fermi pseuodopotential~\cite{fermi1934} or Fano configuration interaction~\cite{fano1961} picture. 
The strong dressing of the $d$ impurity orbitals with electron spins in spin-orbit coupled systems, in a three-atom perturbation framework, could lead to the emergence of DM Hamiltonians~\cite{fret1980,levy1981anisotropy}. The Rydberg $d$ electrons are coupled to all of the ground state ring atoms.

\paragraph*{\bf Acknowledgements.}
S.T.R and H.R.S are grateful to Jim Shaffer for bringing to their attention the work on Friedel bound states.   
S.I.M, S.T.R, and H.R.S acknowledge support from the NSF through a grant for ITAMP at Harvard University. SFY would like to acknowledge funding by the NSF via the CUA PFC and PHY-2207972.

\bibliography{ref}

\onecolumngrid
\newpage
\clearpage
\begin{center}
\large{\bf{Supplemental Material}}
\end{center}
\setcounter{equation}{0}
\setcounter{figure}{0}
\setcounter{table}{0}
\setcounter{page}{1}
\makeatletter
\renewcommand{\theequation}{S\arabic{equation}}
\renewcommand{\thefigure}{S\arabic{figure}}

\section{Derivation of effective interaction Hamiltonian} 

Following the above intuitive explanation for the emergence of DMI, in this section we give a formal derivation of the effective Hamiltonian for a one-dimensional chain of effective spins experiencing dipole-dipole 
 interaction with a moving mediator. 
We consider a setup, illustrated in Fig.\ref{fig:simplified-setup}c, in which $N$ effective spins, encoded in Rydberg states of neutral atoms, with the spin $S=1/2$ 
are 
tightly trapped in a one-dimensional optical lattice or a trap array. A two-level mediator, also encoded in neutral atom Rydberg states, moves with a certain quasimomentum $k_{0}$ 
in another parallel 
one-dimensional optical lattice. 
The many-body Hamiltonian is then $\hat{H}=\hat{H}_{0}+\hat{V}_{dd}$, 
where
\begin{align}
\label{eq:non-int-ham}
\hat{H}_{0}=\sum_{i=1}^{N}E_{\rm spin}\ket{\uparrow_{i}}\bra{\uparrow_{i}}+\sum_{\rm m=g,e}\sum_{k,\nu}{\cal E}_{\rm m}(k_{\nu})\ket{k_{\nu}}\ket{\rm m}\bra{\rm m}\bra{k_{\nu}}, 
\end{align}
represents its the non-interacting part. 
Also, $E_{\rm spin}=E_{\uparrow}-E_{\downarrow}$ is the energy of the spin transition, $\mathcal{E}_{e}(k_{\nu})=E_{\rm m}+E_{\rm kin}(k_{\nu})$ and 
$\mathcal{E}_{g}(k_{\nu})=E_{\rm kin}(k_{\nu})$ include both the 
internal $E_{\rm m}=E_{e}-E_{g}$ and kinetic $E_{\rm kin}(k)$ 
mediator energies, corresponding to its internal states $\ket{g}$ and $\ket{e}$ and a motional state characterized by quasimomentum $k$ in the energy band $\nu$ of the mediator 
lattice.

We further assume that the i-th spin interacts with the mediator via dipole-dipole interaction as in Eq.~(\ref{eq:dip-dip-inter-init}):
\begin{align}
\label{eq:dip-dip-int-second}
\hat{V}=\sum_{i=1}^{N}\sum_{k_\nu,k'_{\nu'}}\ket{k_{\nu}}\bra{k'_{\nu'}}\hat{S}_{i}^{+}\hat{\sigma}_{\rm m}^{-}V^{\uparrow \downarrow,ge}_{k_{\nu},k'_{\nu'}}+\ket{k_{\nu}}\bra{k'_{\nu'}}\hat{S}_{i}^{+}\hat{\sigma}_{\rm m}^{+}V_{k_{\nu},k'_{\nu'}}^{\uparrow \downarrow,eg}+{\rm H.c.},
\end{align}
where the sum is implied to run over all quasimomentum states and bands, while ${\rm H.c.}$ stands for the hermitian conjugate. 
Here, the dipole-dipole interaction matrix element has been integrated over Bloch states, i.e.
\begin{align} 
V_{k_{\nu},k'_{\nu'}}^{ \alpha \beta,\eta \xi}=\bra{ k_{\nu}}\frac{{\bf d}^{\alpha \beta}_{\rm spin}\cdot{\bf d}_{\rm m}^{\eta \xi}}{R_{i \rm m}^{3}}-\frac{3\left({\bf d}^{\alpha \beta}_{\rm spin}\cdot{\bf R}_{i\rm m}\right)\left({\bf d}^{\eta \xi}_{\rm m}\cdot{\bf R}_{i\rm m}\right)}{R_{i\rm m}^{5}}\ket{k'_{\nu'}}.
\end{align}
with $\alpha\beta=\uparrow \downarrow,\downarrow \uparrow$, and $\eta \xi=eg,ge$. We also assume that the diagonal dipole moments of the spin and mediator states are zero, namely $\bra{\uparrow_{i}}\hat{{\bf d}}_{i}\ket{\uparrow_{i}}=\bra{\downarrow_{i}}\hat{{\bf d}}_{i}\ket{\downarrow_{i}}=0$ and 
$\bra{e}\hat{{\bf d}}_{\rm m}\ket{e}=\bra{g}\hat{{\bf d}}_{\rm m}\ket{g}=0$. However, spin and mediator transition dipole matrix elements are non-zero, i.e. $\bra{\uparrow_{i}}\hat{{\bf d}}_{i}\ket{\downarrow_{i}}={\bf d}^{\uparrow \downarrow}_{\rm spin} \neq 0$, and   
 $\bra{e}\hat{{\bf d}}_{\rm m}\ket{g}={\bf d}^{eg}_{\rm m} \neq 0$. 
It is further supposed that the mediator is allowed to move only in 
one dimension, i.e. its motion perpendicular to the lattice is restricted by a deep optical or MW trap.

If the spin-mediator interaction is weak, meaning that $|\hat{V}_{dd}| \ll |E_{\rm spin}-E_{\rm m}|,E_{\rm spin},E_{\rm m}$, 
it induces couplings between many-body spins states of the same energy, corresponding to the same mediator state, similar to the case of two spins depicted in Fig.\ref{fig:simplified-setup}(b). The coupling leads to an effective interaction  between spins and can be obtained using the Schrieffer-Wolff transformation
\begin{align}
e^{\hat{S}}\hat{H}e^{-\hat{S}}=\hat{H}+\left[\hat{S},\hat{H}\right]+\frac{\left[\hat{S},\left[\hat{S},\hat{H}\right]\right]}{2}+O\left(\hat{S}^{3}\right).
\label{eq:Schrieffer-Wolff-expan}
\end{align}
The interaction $\hat{V}_{dd}$ is eliminated by setting $\left[\hat{S},\hat{H}_{0}\right]=-\hat{V}_{dd}$, with the corresponding 
generator 
\begin{align}
\label{eq:gener}
\hat{S}=\sum_{i=1}^{N}\sum_{k_{\nu}k'_{\nu'}}\frac{\ket{k_{\nu}}\bra{k'_{\nu'}}\hat{S}_{i}^{+}\hat{\sigma}_{\rm m}^{-}V^{\uparrow \downarrow,ge}_{k_{\nu},k'_{\nu'}}}{E_{\rm spin}+E_{\rm kin}(k_{\nu})-E_{\rm m}-E_{\rm kin}(k'_{\nu'})}+\frac{\ket{k_{\nu}}\bra{k'_{\nu'}}\hat{S}_{i}^{+}\hat{\sigma}_{\rm m}^{+}V^{\downarrow\uparrow ,eg}_{k_{\nu},k'_{\nu'}}}{E_{\rm spin}+E_{\rm m}+E_{\rm kin}(k_{\nu})-E_{\rm kin}(k'_{\nu'})}-{\rm H.c.}.
\end{align}

The transformed Hamiltonian acquires the form
\begin{eqnarray}
\label{eq:Schrieffer-Wolff-expansion}
e^{\hat{S}}\hat{H}e^{-\hat{S}}=\hat{H}_{0}+\frac{\left[\hat{S},\hat{V}_{dd}\right]}{2}+O\left(|\hat{V}_{dd}|^{3}\right)\approx \hat{H}_{0}+\hat{V}_{\rm eff},
\end{eqnarray} 
in which the effective interactions $\left[\hat{S},\hat{V}_{dd}\right]/2$ is now of the second order in $\hat{V}_{dd}$.  

Combining Eqs.(\ref{eq:dip-dip-int-second}) and (\ref{eq:gener}) and assuming  that initially the mediator is prepared in the ground state $\ket{g}$ and in a motional state $\ket{k_{\nu_0}}$ with a quasimomentum $k_{0}$ in the $\nu_{0}$ Bloch band,   
 the effective interaction
$\hat{{V}}_{\rm eff}=\bra{g,k_{0\;\nu_{0}}}\left[\hat{S},\hat{V}_{dd}\right]/2\ket{g, k_{\nu_{0}}}$ can be obtained in the same form as in Eq.(\ref{eq:Veff-approx}):
\begin{align}
\label{eq:Veff-interm}
\hat{{V}}_{\rm eff}=\sum_{i,j=1;i \neq j}^{N} J_{ij}^{+-} 
\hat{S}_{i}^{+}\hat{S}_{j}^{-}+J_{ij}^{-+}\hat{S}_{i}^{-}\hat{S}_{j}^{+}=\sum_{i,j=1; i \neq j}^{N} \frac{J_{ij}^{\bot}}{2}\left(\hat{S}_{i}^{+}\hat{S}_{j}^{-}+\hat{S}_{i}^{-}\hat{S}_{j}^{+}\right)
+{\bf D}_{ij}\cdot \left(\hat{\bf S}_{i}\times \hat{\bf S}_{j}\right),
\end{align}
with
\begin{align}
\label{eq:J-coeff-main}
J_{ij}^{+-}=\frac{1}{2}\sum_{k_{\nu}}\left(\frac{V_{k_{\nu_0}k_{\nu},i}^{\uparrow \downarrow,ge}V_{k_{\nu}k_{\nu_0},j}^{\downarrow \uparrow,eg}}{E_{\rm spin}+E_{\rm kin}(k_{\nu_{0}})-E_{\rm m}-E_{\rm kin}(k_{\nu})} -\frac{V_{k_{\nu_0}k_{\nu},i}^{\downarrow \uparrow,ge}V_{k_{\nu}k_{\nu_0},j}^{\uparrow \downarrow,eg}}{E_{\rm spin}+E_{\rm m}+E_{\rm kin}(k'_{\nu'})-E_{\rm kin}(k_{\nu_{0}})}\right),
\end{align}
where we have included the $i$ and $j$ subscript to indicate the spins which are coupled by this effective interaction.  We have neglected terms of the order of $\hat{S}_{i}^{\pm}\hat{S}_{j}^{\pm}$ which couple spin states differing in energy by $\pm 2E_{\rm spin}$ by assuming that $|\hat{{\tilde V}}_{\rm eff}| \ll E_{\rm spin}$. The latter condition makes the transitions induced by these terms off-resonant with the underlying probability $\sim|\hat{{\tilde V}}_{\rm eff}|^{2}/E_{\rm spin}$ small.

The effective Hamiltonian of the system 
is indeed of XX type with the additional DMI term, i.e.
\begin{align}
J_{ij}^{\bot}=2ReJ_{ij}^{+-},~~~{\bf D}_{ij}=\left(0,0,2ImJ_{ij}^{+-}\right).
\end{align}
We remark that a more general XXZ 
interaction with independently controllable XX and ZZ parts can also be realized if at least one of the spin states has a non-zero dipole moment.

\begin{figure}[h]
\center{
\includegraphics[width=12.cm]{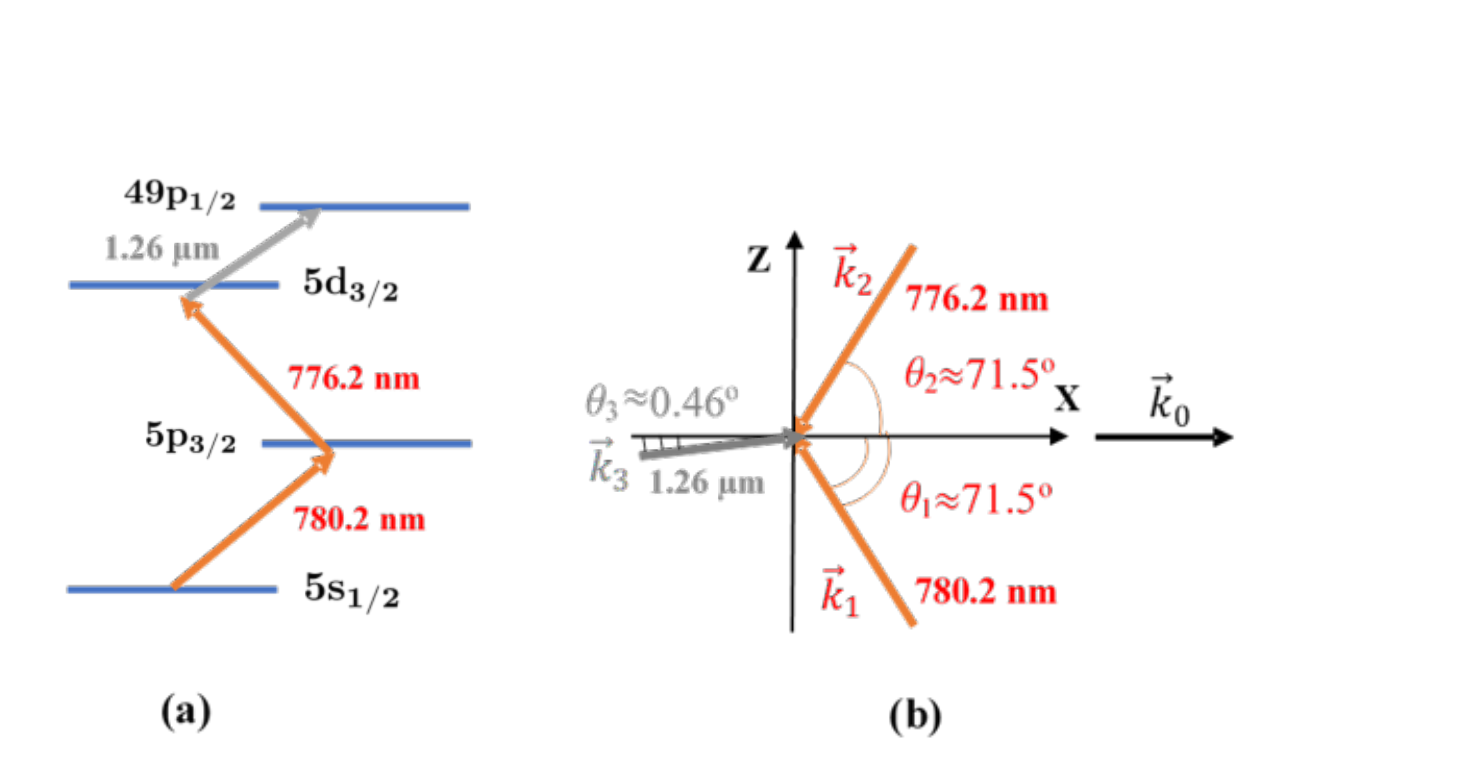}
\caption{\label{fig:K0-angles} {\it Preparation of the mediator moving in free space with the 
momentum ${\bf k}_{0}$ in the $\ket{49p_{1/2}}$ state}. (a) The initial mediator state can be 
prepared via three-photon excitation from the ground $5s_{1/2}$ internal state of Rb via intermediate $5p_{3/2}$ and $5d_{3/2}$ states. The fields 
should be near-resonant with their respective transitions characterized by Rabi frequencies sufficient for excitation and wavelengths  $\lambda_{1} \approx 780.2$ nm, $\lambda_{2} \approx 776.2$ nm and $\lambda_{3} \approx 1.26$ $\mu$m. The fields can be applied in the same 
plane, e.g. the $X-Z$ one, in a star-like geometry to satisfy the condition ${\bf k}_{1}+{\bf k}_{2}+{\bf k}_{3}={\bf k}_{0}$, which can be 
realized with the angles for the wavevectors $\theta_{1} \approx \theta_{2} \approx 71.5^{o}$ and $\theta_{3} \approx 0.46^{0}$.}
}
\end{figure}

\section{Preparation of the initial mediator motional state}

Mediator states with specific $k_{0}$ in the lowest Bloch band can be initialized by first preparing a mediator moving in free space with the 
velocity $v_{0}=\hbar k_{0}/m_{\rm m}$. This can be realized by excitation from the ground $5s_{1/2}$ electronic 
state to the $\ket{g}=\ket{49p_{1/2}}$ state through multiphoton excitation. For example, 
the $5s_{1/2}$ state can be coupled to the $49p_{1/2}$ 
state using a three-photon field via intermediate $5p_{3/2}$ and $5d_{3/2}$ states in a star-like geometry as shown in Fig.~\ref{fig:K0-angles}(a). 
Similar three-photon 
excitation processes were shown to provide recoil-less and Doppler-free excitation of Rb atoms to Rydberg $np$ states~\cite{Doppler-recoiless-Rydb-excit}. 
The energies and the wavevectors of the fields, assumed for 
simplicity to lie in the same plane, should satisfy the conditions $k_{1}+k_{2}+k_{3}=(E_{g}+\hbar^{2}k_{0}^{2}/2m_{\rm m})/c\hbar$ and 
${\bf k}_{1}+{\bf k}_{2}+{\bf k}_{3}={\bf k}_{0}$, 
where ${\bf k}_{1}$, ${\bf k}_{2}$, ${\bf k}_{3}$ 
correspond to the fields coupling to the $5p_{3/2}-5s_{1/2}$, $5d_{3/2}-5p_{3/2}$ and $49p_{1/2}-5d_{3/2}$ transitions, respectively; $E_{g}$ is the energy of the 
$\ket{g}$ mediator state. The frequencies of the fields can be tuned to satisfy the energy constraints, while the angles of the wavevectors can be used to adjust the momentum of the final state. This is achieved here with lasers near resonant with their respective transitions having wavelengths 
$\lambda_{1} \approx 780.2$ nm, $\lambda_{2} \approx 776.2$ nm and $\lambda_{3} \approx 1.26$ $\mu$
while the angles $\theta_{1} \approx \theta_{2} \approx 71.5^{o}$, $\theta_{3} \approx 0.46^{0}$ to be realized, as shown in Fig.\ref{fig:K0-angles}b. The  mediator lattice can be adiabatically ramped up preserving the quasimomentum $k_0$ in the lowest band.

\section{Rydberg states of $^{87}$Rb for spin and mediator encoding}

The splittings of the $48p_{1/2}$ and $49p_{1/2}$ by a magnetic field of magnituide $B$ can be 
calculated using the Breit-Rabi formula, giving the energies of sublevels of a fine-structure doublet in atoms with one outer 
electron~\cite{Atomic-spectroscopy}
\begin{align}
E_{ljm_{j}}(B)=-\frac{\Delta E_{fs}}{2(2l+1)}+\mu_{B}m_{j}B
\pm \frac{\Delta E_{fs}}{2}\sqrt{1+x'^{2}+\frac{4m_{j}x'}{2l+1}},
\end{align}
where $l$, $j=|l\pm 1/2|$ and $m_{j}$ are the orbital and total angular momentum of the atomic state and its projection along the quantization axis. Here, $\pm$ correspond to $j=|l\pm 1/2|$, $x'=(g_{s}-1)\mu_{B}B/\Delta E_{fs}$, $\Delta E_{fs}$ is the fine-structure 
splitting, $\mu_{B}$ is the Bohr's 
magneton, and $g_{s}=2.002319$ is the electron's spin g-factor. The fine-structure splittings for the $49p_{3/2,1/2}$ and $48p_{3/2,1/2}$ states of $^{87}$Rb are $920.971$ MHz and $983.246$ MHz, 
respectively. For $s_{1/2}$ states the Zeeman splitting is given by the expression $g\mu_{B}m_{j}B$ with $g=2$ being the $s_{1/2}$ states g-factor. 
From the above we find that the $m_j=1/2$ states of the $48p_{1/2}-48s_{1/2}$ and $50s_{1/2}-49p_{1/2}$ transistions become resonant at $B\approx 57.55$ G. 
At this magnetic field transitions between other $m_j$ states are detuned by $\sim 50$ MHz.

\section{Implementing DM Hamitonian with spin-orbit Rydberg and ring ground state atoms} \label{IV}

The virtual bound states (vbs) were formulated by Friedel to study the scattering of conduction electrons in metal by localized impurities~\cite{friedel}. The scattering of the electrons by local states of the impurity leads to broadening and shift of the impurity lines, much in the spirit of Fermi pseudopotential scattering or Fano configuration interaction~\cite{fano1961,fermi1934}. 

The indirect spin-spin interaction between local magnetic moments in metals, and nonmagnetic impurity, is ordinarily the isotropic RKKY interaction, i.e. ${\bold{S_i}\cdot \bold{S_j}}$. In the framework of Friedel's vbs, the RKKY effective Hamiltonian has the form,
\begin{equation} 
{\hat V_{\rm eff}= \hat V_{\rm RKKY}} = -\Gamma_i\delta(\bold{r}-\bold{R_i}) {\bold {s}\cdot \bold {S_i}} - \Gamma_j\delta(\bold{r}-\bold{R_j}){\bold {s}\cdot \bold {S_j}},
\end{equation}
where the short-range coupling parameters, $\Gamma$, are proportional to the interaction of the conduction electrons with the localized spins, while $r$ and $R$ refer, respectively, to the electron and the localized spin coordinates.

The Fermi pseudopotential electron-spin interaction delocalizes the Rydberg $d$ orbital allowing for the Rydberg atom to mediate spin-spin interactions as schematically shown in Fig.~\ref{fig:ring}. 
The splitting between different large magnetic quantum number states due to spin-orbit coupling provides the necessary symmetry breaking. 
The effective DM Hamiltonian, in the lowest-order correction to the ground-state energy due to the three-atom perturbation, reads as~\cite{fret1980,levy1981anisotropy}
\begin{eqnarray}
\label{eq:DMI}
    \hat V_{\rm eff}=\hat V_{\rm DMI}  \propto   {(\bold{R_i} \cdot \bold{R_j})} ({ \bold{R_i} \times \bold{R_j})} \cdot ({\bold{ S_i} \times \bold{S_j})} 
      \sim  \bold{D_{ij}} \cdot ({\bold{ S_i} \times \bold{S_j})}.
\end{eqnarray}
It is now clear how the DMI vector is comprised of the dot and cross products of the distance vectors connecting the spin sites to the imourity site. 

The DM Hamiltonian can be simulated with Rydberg spin-orbit interaction in a ring geometry, illustrated in Fig.~\ref{fig:ring}. In this context, a central Rydberg atom excited into a $d$ orbital (Cs($nd$)), interacts via the Fermi pseudopotential~\cite{fey2019} with all ground state atoms in the ring.

\begin{figure}
\center{\includegraphics[width=6.cm]{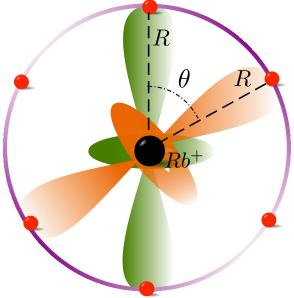}
\caption{\label{fig:ring}Ring geometry for realizing the DM Hamiltonian with anisotropic spin-orbit interactions in Rydberg $d$ orbitals. 
Excitations into high $nd$ states, such as Cs($n> 40d$) would create interactions between a Rydberg- and all ground-state- atoms in micron size rings~\cite{fey2019}, necessary for generating effective DM interactions in Eq.~(\ref{eq:DMI}).} }
\end{figure}

The simplest manifestation is one in which the Rydberg electron at position ${r}$ and spin ${s}$ interacts with two ground state alkali metal atoms at positions ${\bf R}_{i,j}$ and spins $\vec S_{i,j}$. Within the low-energy $s$-wave scattering framework, the interaction Hamiltonian is $\hat{V}=\hat{V}_1+ \hat{V}_2$ ~\cite{fey2019}, where
\begin{align}
\begin{split}
\hat{V}_{i}& =2 \pi  \delta ({r}-{R}_{i}) \left[ a^S_s\hat{P}^S_{i} + a^T_s \hat{P}^T_{i}\right].
\label{eqn:pseudopotential}
\end{split}
\end{align}
The superscripts $\hat{P}^S$ and $\hat{P}^T$ are the operators projecting into the total electron spin singlet and triplet states, i.e. $\hat{P}^T_{i}= {\bf s}\cdot {\bf s}_{i} +3/4 $ and $\hat{P}^S_{i}=1 -\hat{P}^T_{i} $. Due to its large fine and hyperfine structure splittings, ${}^{133}$Cs is a good candidate for achieving the proposed scheme\cite{fey2019,markson2016}. By adjusting the size of the ring of ground state atoms to coincide with the classical turning point of the Rydberg electron in the core-electron potential, the effect of the s-wave interaction can be maximized.  The influence of the p-wave electron-atom scattering in Cs can be safely mitigated, by choosing excitations with  principal quantum numbers, $n\gtrsim 40$, where the p-wave resonances are pushed to smaller electron-core distances~\cite{tallant2012}. The relativistic spin interactions and the angle-($\theta$) dependence of the three-atom potential energy surface, induce at the impurity site, strong spin-orbit coupling of the $d$ orbitals, resulting in large shifts of the Rydberg lines.  The strong admixture is reminiscent  of the vbs model with spin-orbit interaction~\cite{fret1980,levy1981anisotropy} which leads to the realization of the DM Hamiltonian in Eq.~\ref{eq:DMI}.

\end{document}